\documentclass[%
 reprint,
 amsmath,amssymb,
 aps,
]{revtex4-2}

\usepackage{graphicx}
\usepackage{dcolumn}
\usepackage{bm}
\usepackage{braket}
\usepackage{comment}
\usepackage{bm}

\newcommand{\Eq}[1]{Eq.\,(\ref{#1})}
\newcommand{\Eqs}[2]{Eqs.\,(\ref{#1}) and (\ref{#2})}
\newcommand{\Eqsss}[3]{Eqs.\,(\ref{#1}), (\ref{#2}), and (\ref{#3})}

\newcommand{\Fig}[1]{Fig.\,\ref{#1}}

\newcommand{\Sec}[1]{Sec.\,\ref{#1}}

\newcommand{\be}{\begin{equation}}
\newcommand{\ee}{\end{equation}}
\newcommand{\bea}{\begin{eqnarray}}
\newcommand{\eea}{\end{eqnarray}}
\newcommand{\omtelow}{\Omega^{\text{TE}}_{\text{low}}}
\newcommand{\omtmup}{\Omega^\text{TM}_\text{up}}
\newcommand{\omteup}{\Omega^\text{TE}_\text{up}}
\begin{document}

\preprint{APS/123-QED}

\title{Extended frequency range of transverse-electric surface plasmon polaritons  in graphene}

\author{Zeeshan Ahmad}
\author{Egor A. Muljarov}
\author{Sang Soon Oh} \email{OhS2@cardiff.ac.uk}
\affiliation{ School of Physics and Astronomy, Cardiff University, Cardiff CF24 3AA, United Kingdom }

\date{\today}

\begin{abstract}
The dispersion relation of surface plasmon polaritons in graphene that includes optical losses is often obtained for complex wave vectors while the frequencies are assumed to be real.
This approach, however, is not suitable for describing the temporal dynamics of optical excitations and the spectral properties of graphene.
Here, we propose an alternative approach that calculates the dispersion relation in the complex frequency and real wave vector space. This approach provides a clearer insight into the optical properties of a graphene layer and allows us to find the surface plasmon modes of a graphene sheet in the full frequency range, thus removing the earlier reported limitation ($1.667<\hbar\omega/\mu<2$) for the transverse-electric mode.
We further develop a simple analytic approximation which accurately describes the dispersion of the surface plasmon polariton modes in graphene. Using this approximation, we show that
transverse-electric surface plasmon polaritons propagate along the graphene sheet without losses even at finite temperature.

\end{abstract}

\maketitle

\section{\label{sec:level1}Introduction}
Surface plasmon polaritons (SPPs) are collective excitations of charge density coupled to electromagnetic waves that can travel along a conductor-dielectric interface~\cite{Raether1988}.
Interestingly, even an atomically thin conducting layer, such as a graphene sheet, can support SPPs~\cite{Koppens2011}.
In a graphite intercalated compound that contains multiple non-interacting two-dimensional graphene layers, SPPs were modelled theoretically and observed experimentally~\cite{Shung1986}.
The first discovery of graphene~\cite{Novoselov2004} triggered more theoretical studies on SPPs in this material~\cite{Vafek2006,Hanson2008, Jablan2009}.
Since the first experimental observation of SPPs in graphene~\cite{Chen2012}, more efforts have been made to use the SPPs for controlling the optical properties of graphene, such as electronically tuned extraordinary transmission~\cite{Kim2016}, mode confinement by gap plasmons~\cite{Francescato2013}, and resonant absorption by an antidot array~\cite{Nikitin2012}.

Although most studies of SPPs have focused on transverse-magnetic (TM) polarization, a single graphene layer, unlike normal-metal sheets, can also support transverse-electric (TE) SPP modes~\cite{Mikhailov2007}.
Technically, the existence of each type of SPP modes can be confirmed by solving a dispersion relation
between the frequency and the wave vector following from Maxwell's equations and the conductivity model.
For instance, the SPP dispersion for a dielectic/metal interface or a thin film of a Drude metal allows only TM modes which have zero magnetic field components normal to the interface and along the propagation direction.
By analyzing the dispersion relation for complex wave numbers and real frequencies, Mikhailov {\it et al.} have shown~\cite{Mikhailov2007} that TE SPP modes in graphene exist only in the range $1.667<\hbar\omega/\mu<2$ at zero temperature, where $\omega $ and $\mu$ are, respectively, the light frequency and the chemical potential. Here, the lower limit corresponds to the zero of the imaginary part of the optical conductivity, while the upper limit is given by the minimum of the interband transition energy at zero temperature.


Importantly, the lower limit ($\approx$1.667) for the normalized frequency $\hbar\omega/\mu$ of the TE SPP mode  can vary because the imaginary part of the optical conductivity of graphene may change depending on temperature and gate voltage.
In addition, the range for the TE SPP mode frequency $\omega$ itself can be tuned by changing the chemical potential $\mu$ which in turn may be controlled by applying gate voltage or external magnetic field~\cite{He2013}.
In the literature, the TE SPP mode solution has been studied for a graphene layer sandwiched between two dielectric media, and it was found that the range of the TE SPP mode frequencies can be modified by changing the permittivity contrast~\cite{Kotov2013}.
Furthermore, the range of TE SPP modes can be reversed  when the surrounding material has a negative refractive index~\cite{Zhang2020}.
TE SPP modes in a graphene sheet placed on top of a nonlinear material substrate were also found to be limited to a similar frequency range~\cite{Bludov2014}.

In this work, we focus on the dispersion of SPP modes in a graphene layer with finite temperature and non-zero chemical potential and show that a complex-frequency analysis developed in this paper removes both the upper and the lower limits for the TE SPP mode in graphene.

While it is well-known how to calculate the dispersion relations of SPPs in graphene for given optical conductivities~\cite{Mikhailov2007}, this has been done assuming that any SPP mode has a real frequency but complex wave number $q = q'+iq''$~\cite{Bludov2014,Mikhailov2007,Falkovsky2007}.
This corresponds to a continuous-wave excitation of a SPP which has a finite propagation length within the graphene layer of the order of $1/q''$. This picture is more suited for describing electromagnetic waves propagating in inhomogeneous waveguides not conserving the in-plane component of the wave number $q$.
In contrast, uniform waveguides conserve $q$ which can naturally be taken real. In our approach, assuming the excitation of the system is limited in time, the temporal evolution of SPP modes is described by a complex frequency $\omega=\omega'+i\omega''$, with typically $\omega''<0$ corresponding to a temporal decay due to radiative losses or absorption. The main advantage of this approach is that it provides a direct access to the optical spectra of the system where the imaginary parts of complex frequencies of isolated modes usually correspond to the half width at half maximum of the resonance peaks.
A. Page {\it et al.}~\cite{Page2015,Page2018} have recently shown that the complex frequency approach can be used to describe optical gain (for $\omega''>0$) in a non-equilibrium inverted graphene system where TM SPP modes were considered.

The complex-frequency approach has recently become widespread and broadly used in optics, owing to the
useful concept of resonant states~\cite{Weinstein69,MuljarovEPL10}, also known in the literature as quasi-normal modes~\cite{Lalanne18}. These are the eigen solutions of Maxwell's equations satisfying outgoing wave boundary conditions. They present a rigourous and powerful tool for analyzing optical spectra, such as scattering and transmission~\cite{LobanovPRA18,WeissPRB18}, with the real part of the complex frequency of the resonant state typically corresponding to the frequency position of a spectral line and the imaginary part to the half of its linewidth. Physically, resonant states form as a result of constructive interference of multiply reflected electromagnetic waves from the boundaries or inhomogeneities within optical systems. They have been studied in the literature both in finite optical systems, such as dielectric~\cite{DoostPRA14,RybinPRL17,LobanovPRA19} and plasmonic nanoparticles~\cite{SauvanPRL13,MuljarovPRB16,SehmiPRB20}, and in infinitely extended systems, such as planar waveguides~\cite{ArmitagePRA14} and photonic crystals~\cite{TikhodeevPRB02,WeissPRL16,NealePRB20}. Although in atomically thin films, such as a single graphene layer, these states do not normally exist, the formalism of complex-frequency modes can still be very useful, as we show in this paper.

The paper is organized as follows. Section~\ref{Sec:cond} introduces the optical conductivity of a single graphene layer. Section~\ref{Sec:secular} describes the secular equations determining the dispersion of SPPs in TM and TE polarizations. Sections~\ref{Sec:modes} and \ref{Sec:threshold} present the main results of the paper, including both exact numerical and approximate analytical solutions of the secular equations in the complex frequency plane, their dependence on the propagation constant, temperature, and the chemical potential, and elimination of both the lower and the upper boundaries for the TE SPP mode frequencies. The temperature dependence of the threshold frequencies is discussed in Sec.\,\ref{Sec:temperature}. Section~\ref{Sec:summary}~summarizes the results of the paper. Appendices~\ref{App:Intraband}--\ref{Sec:gamma} provide details on derivations of the optical conductivity of graphene and secular equations for the TM and TE SPP modes, and supply an additional material on our study of the TE mode near the lower threshold frequency and on the SPP dispersion at a finite damping.


\section{Optical conductivity of graphene }
\label{Sec:cond}
The conductivity of graphene has an ``interband" term in addition to the usual metallic Drude term, also referred to as ``intraband". It is the interband term of the conductivity which gives rise to TE SPP modes found in a graphene layer in contrast to a Drude-metal layer where such modes do not exist.
Following Refs.~\cite{Gusynin2006,Falkovsky2007,Mikhailov2007}, we derive in Appendices~\ref{App:Intraband} and \ref{App:Interband} the two-dimensional (2D) optical conductivity of a homogeneous graphene sheet. In the long-wavelength limit, the expression for the 2D optical conductivity is given as a function of the light frequency $\omega$ by
\begin{widetext}
\be
    \sigma(\omega) = i \alpha \left\{  \frac{2\ln{(2+2\cosh{\mu\beta})}}{\mu\beta(\Omega+i\Gamma)}
    +\int_0^\infty dE \left[ N(-E;\mu\beta)-N(E;\mu\beta)\right]\left(\frac{1}{\Omega+i\Gamma-2E}+\frac{1}{\Omega+i\Gamma+2E}\right)\right\}\,,
    \label{sigma_formula}
\ee
\end{widetext}
where
\be
N(E;\xi)=\frac{1}{e^{\xi(E-1)}+1}
\label{distr}
\ee
is the Fermi-Dirac distribution,  $\Omega = \hbar\omega/\mu$ is the normalized frequency,  $\beta =(k_BT)^{-1}$ is the inverse temperature, $\alpha=e^2/(\hbar c)$ is the fine-structure constant, and $\Gamma$ is a phenomenological damping. The first term in the curly bracket in \Eq{sigma_formula} arises from intraband transitions, whereas the second one comes from interband transitions near the Dirac point in graphene dispersion. Importantly, \Eq{sigma_formula} has an analytic dependence on $\Omega$ which can be continued into the complex $\Omega$-plane without changing the formula. Note that for real
$\Omega$ and $\Gamma=0$, the integrand encounters a pole on the integration path. This requires that the diverging integral  is split into a principal-value part and a half-pole contribution that can technically be achieved by keeping $\Gamma$ positive infinitesimal in \Eq{sigma_formula}.
For $\text{Im}\,\Omega\neq 0$, this is no longer needed. However, the interband term is represented by a multi-valued function having a logarithmic nature. Therefore, for the analytic continuation, one has to choose the right Riemann sheet which provides the proper values of the integral at real $\Omega$.

\begin{figure}[htp]
\centering
\includegraphics[width=\columnwidth]{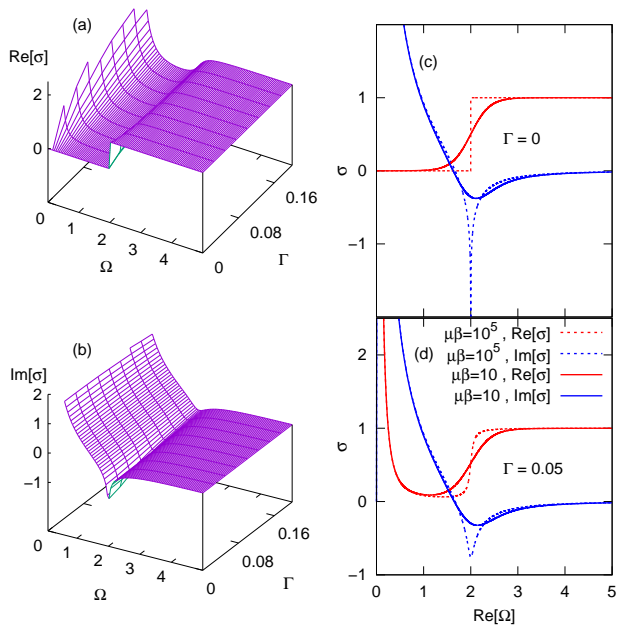}
\caption{Surface plots of (a) the real part $\sigma'=\text{Re}\,\sigma$ and (b) the imaginary part $\sigma''=\text{Im}\,\sigma$ of the 2D optical conductivity of graphene $\sigma$  shown in units of the fine-structure constant $\alpha$ for near zero temperature ($\mu\beta = 10^5$) as functions of normalized real frequency $\Omega$ and damping $\Gamma$. Dependence of $\sigma'$ and $\sigma''$ on frequency only (c) without and (d) with damping, for $\mu\beta = 10^5$ and $\mu\beta = 10$.}
\label{fig:condcomplex}
\end{figure}

Figures~\ref{fig:condcomplex}\,(a) and (b) show the graphene conductivity as a function of normalized real frequency $\Omega=\hbar\omega/\mu$ and damping $\Gamma$ at zero temperature. Increasing temperature smoothes out the Fermi distribution which is reflected by the smearing the step-like feature in the real part and the logarithmic divergence in the imaginary part of the interband conductivity at $\Omega=2$. The intraband Drude-like component of \Eq{sigma_formula} has a zero-frequency  pole which is moving away from the real axis as the damping increases. Figures~\ref{fig:condcomplex}\,(c) and (d) show cross-sections of the surface plots at fixed values of $\Gamma$ and finite ($\mu\beta = 10$) and almost zero ($\mu\beta = 10^5$) temperatures.
Due to the functional dependence of the conductivity \Eq{sigma_formula} on $\Omega+i\Gamma$, \Fig{fig:condcomplex} can also be understood as complex-frequency plots of the conductivity, treating $\Omega$ as the real and $\Gamma$ (or its portion) as the imaginary part of frequency.

In the following we will be using, where it is convenient, the frequency $\Omega$, wave number $Q$, and temperature $(\mu\beta)^{-1}$ normalized with respect to the chemical potential $\mu$, treating the latter as a natural scaling parameter for, respectively, the frequency $\omega$, wave number $q$, and temperature $T$. The chemical potential itself can be controlled e.g. by charge carrier concentration of graphene~\cite{Falkovsky2008} and is also linked to the damping $\Gamma$. In the absence of voltage or in an undoped graphene layer, the valence band is fully filled and the conduction band is empty at zero temperature, implying that $\mu=0$ and $\Gamma=0$. The concentration of free carriers -- electrons in the conduction band and holes in the valence band -- is zero, $n_0=0$. In a doped graphene layer or in the presence of voltage, the concentration of free carriers $n_0$ determines the chemical potential via the following equation~\cite{Falkovsky2008, Falkovsky2008a}:
\begin{align}
    n_0=\frac{2\mu^2}{\pi(\hbar V)^2}\int_0^\infty [N(E;\mu\beta)-N(E+2;\mu\beta)] EdE\,,
\label{conc}
\end{align}
where $V$ is the electron Fermi velocity in graphene. At zero temperature, the above integral is equal to 1/2, which gives $n_0= \mu^2/(\pi\hbar^2 V^2)$. At a non-zero temperature, this integral depends on $\mu$ and $T$, in accordance with \Eq{distr}, thus introducing a temperature-dependent correction to the above expression for $n_0$. The damping $\Gamma$ contributing to \Eq{sigma_formula} increases with density of impurities~\cite{Falkovsky2007}.

\section{SPP modes in graphene}
\subsection{Secular equations for SPP modes}
\label{Sec:secular}

In this subsection, we present the secular equations determining in each polarization the dispersion relation between the SPP mode frequency $\omega$ and the propagation wave number $q$. We briefly describe their derivation for an infinitely thin graphene sheet. To gain physical insight of SPP modes in graphene, we discuss, at the end of this subsection, a comparison between SPP modes in graphene and in a Drude metal.

For a very thin planar conducting layer, the dispersion relation of SPP modes can be obtained in two ways. One may start deriving from Maxwell's equations and boundary conditions a secular equation for the SPP modes for a finite-thickness conducting material with a bulk conductivity. This secular equation can then be simplified in the limit of an infinitesimal film thickness. Alternatively, one may obtain the secular equation and SPP dispersion relation by assuming an infinitesimal layer with a surface conductivity $\sigma(\omega)$. The two ways lead to  identical results. We follow the second approach and derive in Appendix~\ref{App:modes} the secular equations for both polarizations, assuming the conducting layer is placed at $z=0$ and is surrounded by vacuum, which is expressed by the permittivity in the entire space
\be
\varepsilon(\omega;z)=1+ \frac{2i\sigma(\omega)}{\omega}\,\delta(z)\,,
\ee
where  $\delta(z)$ is the Dirac delta function.
Using for brevity the units in which the speed of light in vacuum $c=1$, the secular equations for SPP modes in both polarizations are given by
\begin{align}
    \omega +  k(\omega)\sigma(\omega)&=0\quad \text{ (TM)}\,,
    \label{tm_secular}\\
    k(\omega)+\omega \sigma(\omega)&=0\quad \text{ (TE)}\,,
    \label{te_secular}
\end{align} where $k(\omega)=\sqrt{\omega^2-q^2}$ and $q$ are, respectively, the normal and in-plane components of the light wave vector in vacuum, and $\omega$ is the complex light frequency.
The dimensionless surface conductivity $\sigma(\omega)$ of a graphene layer to be used  in the above equations is given by \Eq{sigma_formula}. Equations~(\ref{tm_secular}) and (\ref{te_secular})
determine the dispersion relations between the real in-plane wave number $q$ and complex frequency $\omega$.
Note that the same equations were used in the literature~\cite{Falko1989,Mikhailov2007} for finding SPP modes at a fixed real frequency of light, thus determining instead from \Eqs{tm_secular}{te_secular} complex propagation constants of SPPs.

Now, before moving on to graphene, we briefly describe the solution of the secular equations~(\ref{tm_secular}) and (\ref{te_secular}) for a Drude metallic sheet. In this case, the conductivity would only consist of the first term in \Eq{sigma_formula}, so that $\sigma(\omega)\propto i/\omega$, and \Eq{tm_secular} results in a TM mode having a square-root dispersion, $\omega\propto \sqrt{q}$. We find in the next subsection a similar SPP mode for graphene in the frequency range dominated by the Drude conductivity. It can be seen that \Eq{te_secular} for TE polarization has no solution with a non-zero real part of frequency for the Drude conductivity. We found, however, a TE mode of a Drude metallic sheet which has a purely imaginary frequency: $\text{Re}\,\omega=0$. This TE mode is also purely decaying in time: $\text{Im}\,\omega<0$. However, with the full graphene conductivity, there is also a propagating SPP mode in TE polarization~\cite{Mikhailov2007}, with $\text{Re}\,\omega\neq 0$, which we discuss in more depth in the rest of this section.

\subsection{SPP dispersion: Exact results and analytic approximations}
\label{Sec:modes}

In this subsection we present graphene SPP dispersion for both TM and TE modes which are found by solving the secular equations~(\ref{tm_secular}) and (\ref{te_secular}) numerically and comparing with a developed analytic approximation. We show in particular that the mode spectral linewidth (given by $\text{Im}\,\omega$) is controlled by the temperature $\mu\beta$ contributing via the Fermi function \Eq{distr}. We analyze the earlier reported~\cite{Mikhailov2007} limited frequency range $1.667<\hbar\omega/\mu<2$ for the TE mode and remove both the upper and the lower boundaries for this mode.
\begin{figure}[t]
\centering
\includegraphics[width=\columnwidth]{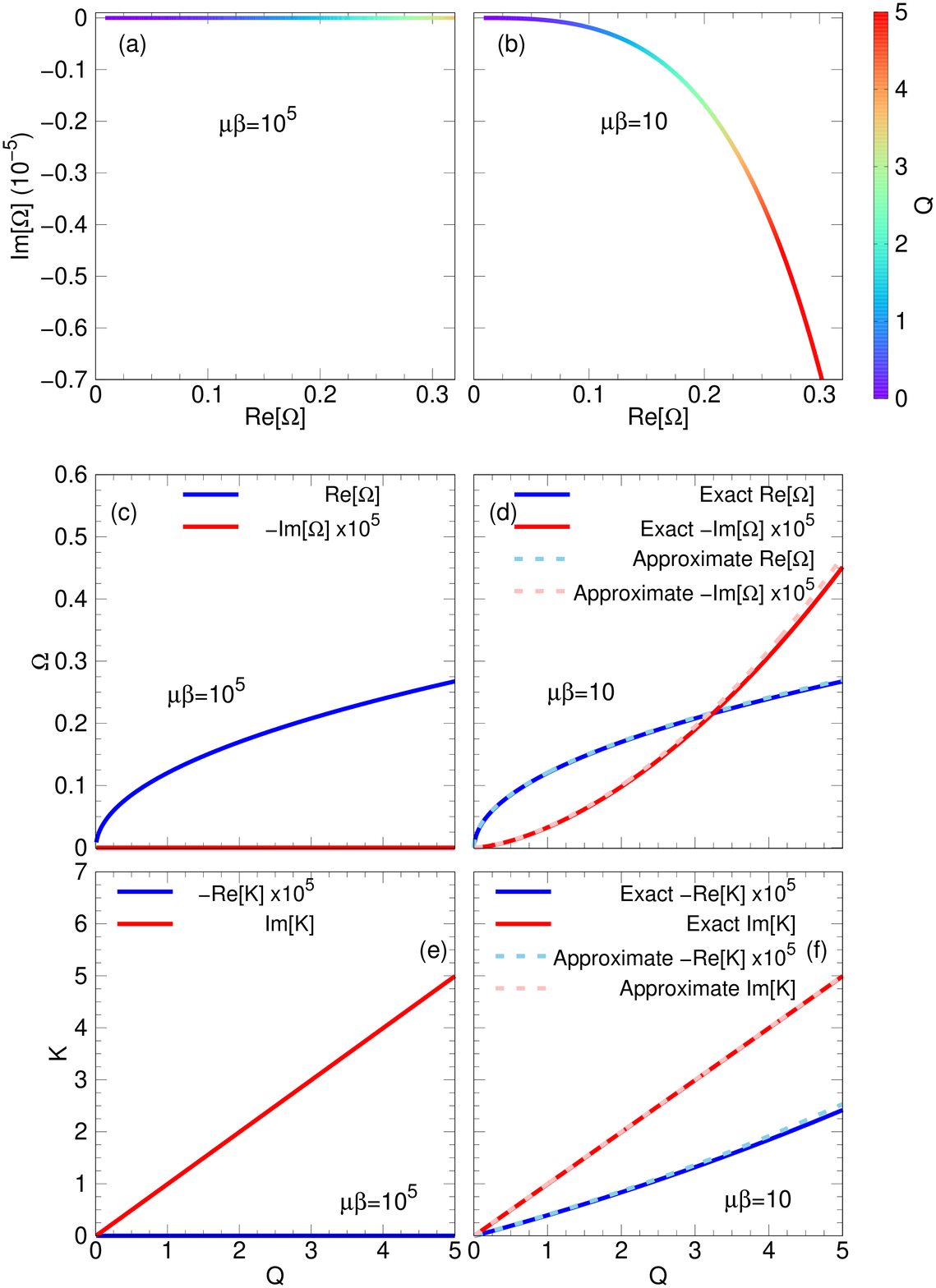}
\caption{
(a,b) Complex eigenfrequency $\Omega=\hbar\omega/\mu$ of TM surface plasmon mode of a graphene layer shown in the complex frequency plane as function of the real in-plane wave number of light $Q=\hbar qc/\mu$ given by the colour code. Real and imaginary parts of (c,d) the eigenfrequency $\Omega$ and (e,f) the normal component of the light wave number $K=\hbar kc/\mu$ calculated by solving \Eq{tm_secular} exactly (solid lines) and using the approximation \Eqs{w_TM}{k_TM} (dashed lines). The data is presented for (a,c,e) high temperature ($\mu\beta = 10$) and (b,d,f) low temperature ($\mu\beta = 10^5$), and $\Gamma=0$.} \label{tm_plane}
\end{figure}

\begin{figure}[t]
\centering
\includegraphics[width=\columnwidth]{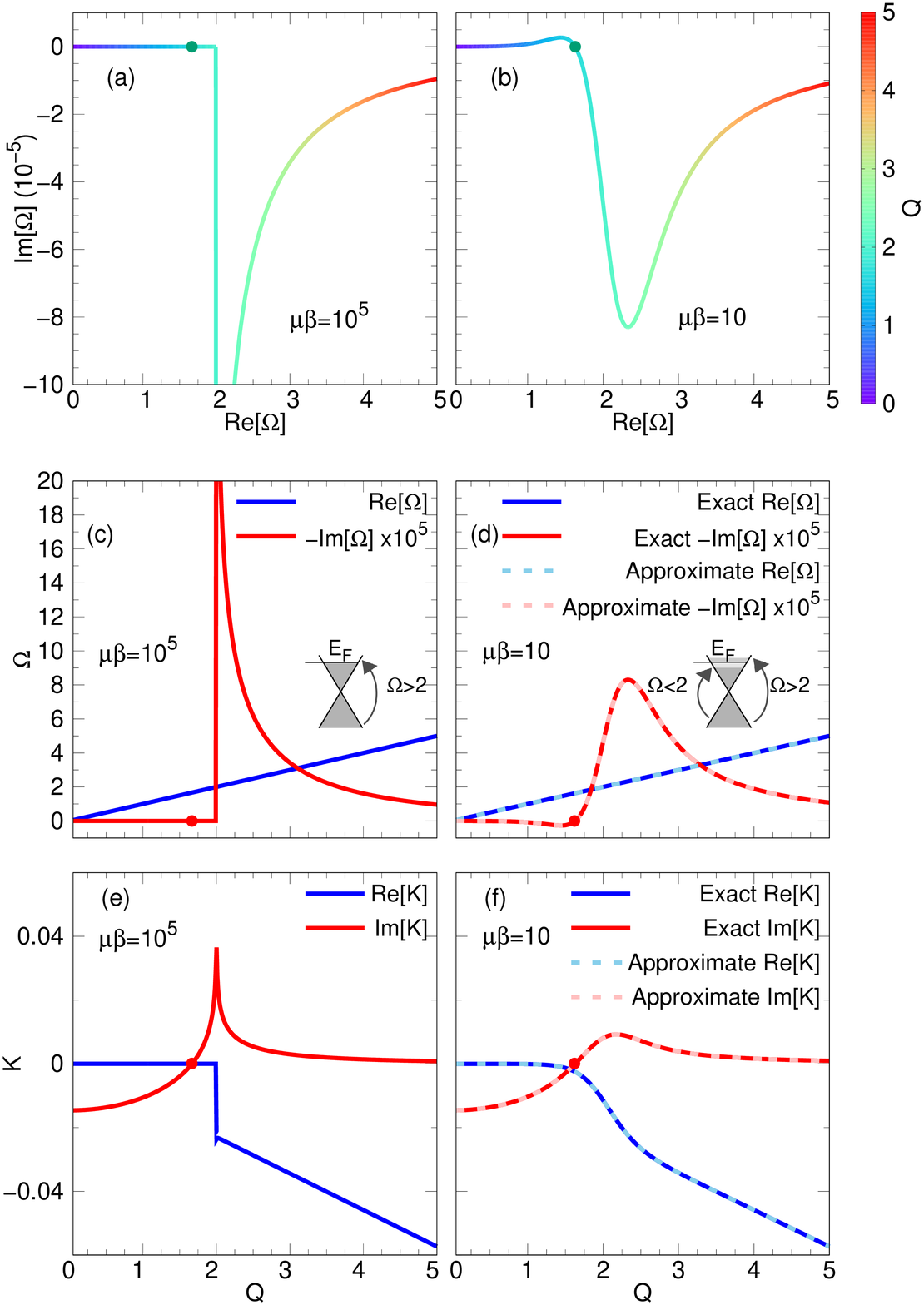}
\caption{
As \Fig{tm_plane} but for the TE mode determined by \Eq{te_secular} and the approximation given by \Eqs{w_TE}{k_TE}. Blue and red dots show the position of the lower threshold $\omtelow$.
Insets: sketches of the electronic dispersion in graphene.
}
\label{te_plane}
\end{figure}

Figure~\ref{tm_plane} shows the dispersion of the TM mode, which includes both the complex $\omega$ plots and dependencies on $q$ of the real and imaginary parts of $\omega$ and $k$, all shown at high ($\mu\beta=10$) and low ($\mu\beta=10^5$) temperatures. The TM dispersion lies well below the light line, $\omega=q$, as can be seen from Figs.\,\ref{tm_plane}\,(c) and (d) by comparing magnitudes of $\text{Re}\,\omega$ and $q$. In this frequency regime, the conductivity is dominated by intraband transitions within graphene band structure, described by the Drude-like term in \Eq{sigma_formula}. At low temperature, both $\text{Im}\,\omega$ and $\text{Re}\,k$ are exponentially small, i.e. proportional to $\exp(-\mu\beta)$. Note that $\text{Re}\,\omega$ asymptotically approaches the temperature dependent limit $\omtmup$ for large $q$.
The magnitude of $\text{Im}\,\omega$ increases swiftly with $q$, with the rate determined by temperature, as more charge carrier vacancies become available below the Fermi level, compare Figs.\,\ref{tm_plane}\,(a) and (b). The normal component of the wave number $k$ shown in Figs.\,\ref{tm_plane}\,(e) and (f) demonstrates that the SPP mode is localized in the $z$ direction, due to $\text{Im}\,k\approx q>0$. At the same time, since $\text{Re}\,k<0$ there is a propagation of light towards the conducting sheet, although it is much smaller than $q$.

We also show in Figs.\,\ref{tm_plane}\,(d) and (f), by dashed linesm the following approximate relations derived in Appendix~\ref{tm_appendix}:
\bea
\omega(q)\approx&\sqrt{\omega_0q}-iq\sigma'/2\,,
\label{w_TM}
\\
k(q)\approx&-\sqrt{\omega_0q}\,\sigma'/2+iq\,,
\label{k_TM}
\eea
where $\sigma'$ is the real part of the conductivity (taken at the mode frequency, $\omega\approx\sqrt{\omega_0q}$). The approximation is based on the fact that in this frequency range, the conductivity is dominated by the Drude term, so that its imaginary part is given by $\sigma'' \approx \omega_0/\omega$ whereas the real part can be treated as a small correction, i.e. $|\sigma'|\ll|\sigma''|$.

Figure~\ref{te_plane} presents a dispersion of the TE mode in graphene, again showing both the complex $\omega$ plots and dependencies $\omega(q)$ and $k(q)$. The SPP mode in this polarization of light is unique to the graphene conductivity~\cite{Mikhailov2007} due to the interband part of \Eq{sigma_formula} and does not exist in a normal (e.g. Drude) metallic sheet. The real part of the dispersion curve, $\text{Re}\,\omega$, lies close to the light line $\omega=q$, while the imaginary part, $\text{Im}\,\omega$, is a few orders of magnitude smaller than the real part, as it is clear from Figs.\,\ref{te_plane}\,(c) and (d). Interestingly, both the real and imaginary parts of $k$, albeit being small compared to $q$, are now comparable to each other, in contrast to the TM SPP mode  (compare Figs.\,\ref{tm_plane}\,(f) and \ref{te_plane}\,(f)).

An approximate analytic solution was also obtained for the TE mode, by using the fact that away from the $\omega=0$ pole, the graphene conductivity is small, $|\sigma(\omega)|\ll 1$, as it is proportional to the fine-structure constant which is a small number, $\alpha\ll 1$. In this limit, the solution to \Eq{te_secular} takes the form:
\bea
\omega(q)\approx&q+iq\sigma'\sigma''\,,
\label{w_TE}
\\
k(q)\approx&-q\sigma'-iq\sigma''\,,
\label{k_TE}
\eea
where $\sigma'$ and $\sigma''$ are both taken at $\omega=q$, see Appendix~\ref{te_appendix} for derivation. These complex frequency $\omega(q)$ and wave number $k(q)$ are plotted as dashed curves in Figs.\,\ref{te_plane}\,(d) and (f), showing excellent agreement with the exact solution.

For completeness, we also show in  Appendix~\ref{Sec:gamma} the SPP mode dispersion both in TM and TE polarizations for the graphene conductivity at a non-zero damping of $\Gamma=0.05$.

\subsection{Removing the boundaries for the TE mode}
\label{Sec:threshold}

The TE mode dispersion in graphene has been studied in Ref.~\cite{Mikhailov2007} at zero temperature, with the mode frequency reaching but never exceeding the upper boundary at $\omteup=2$. Our complex-frequency analysis allows us to eliminate this boundary, even for low temperatures.
In fact, we see from \Fig{te_plane} that the TE mode exists both below (Re\,$\Omega<\omteup$) and above (Re\,$\Omega>\omteup$) the threshold. Furthermore, we observe, by comparing Figs.\,\ref{te_plane}\,(c) and (d), that the temporal loss, which is given by $-\text{Im}\,\Omega$, increases (decreases) with temperature below (above) the threshold. This can be understood simply as a smearing effect of the electronic distribution over the graphene band structure as temperature rises. Note that the threshold frequency is exactly twice the Fermi level  $E_F$ of graphene, see the insets in Figs.\,\ref{te_plane}\,(c) and (d) which provide sketches of the electronic dispersion. At zero temperature, interband absorption only takes place above $\omteup$ since no charge carriers occupy the electronic bands above the Fermi level, and so the losses are high for Re\,$\Omega>\omteup$ and zero for Re\,$\Omega<\omteup$. When temperature is finite, some charge carriers occupy energy states just above the Fermi level, thus the interband absorption decreases for Re\,$\Omega>\omteup$, thus reducing the losses. At the same time, vacancies of charge carriers are formed below the Fermi level at finite temperature and interband absorption can take place also for Re\,$\Omega<\omteup$, so for this region the losses increase with temperature. This increase of losses is also reflected by the real part of the conductivity. In fact, it is clear from \Fig{fig:condcomplex}\,(c) that the real part is smeared around Re\,$\Omega=\omteup$ as the temperature increases. No similar effects are observed for the TM mode.

Figure~\ref{te_plane} also demonstrates the earlier reported in the literature~\cite{Mikhailov2007} lower threshold $\omtelow$ for the TE mode, which was observed at $\omtelow\approx1.667$ at zero $T$. It is shown in Figure~\ref{te_plane} by blue dots in the complex frequency plane and by red dots on the imaginary parts of the mode frequency $\omega$ and the normal component of the wave number $k$.
Both imaginary parts change their sign at this threshold due to the change of sign of $\sigma''$ [see \Eqs{w_TE}{k_TE} and also \Fig{fig:condcomplex}\,(c)]. Physically, this threshold frequency corresponds to a condition that the intraband and interband electronic transitions are in balance. From a technical viewpoint, however, a positive imaginary part of the frequency, $\text{Im}\,\omega>0$, observed below the threshold, implies an exponential growth in time of the electric and magnetic fields (at any given point in space), whereas a negative imaginary part of the wave number, $\text{Re}\,k<0$, also observed below the threshold, means an exponential growth of the field in space away from the graphene layer. While the latter is typical for radiative modes~\cite{MuljarovEPL10} and thus seems acceptable, the former usually corresponds to a gain~\cite{SehmiPRB20}, which is obviously not present in this system. One could therefore conclude that the TE mode does not exist below the threshold. However, the complex-frequency analysis allows us again to understand the properties of the SPP mode near the threshold and to eliminate this lower boundary.

To see that the TE mode exists both above and below the lower threshold at $\omtelow$, we consider the spatial and temporal behaviour of the electric field together, having the following explicit form $E(x,z;t)=E_0e^{i(qx+kz-\omega t)}$, where $E_0$ is a constant, and we have taken $z>0$ for definiteness (see Appendix~\ref{App:modes} for the analytic form of the fields). Now, separating the real and imaginary parts of the frequency, $\omega=\omega'+i\omega''$, and of the normal component of the wave number, $k=k'+ik''$, and using the fact that the in-plane component of the wave number $q$ is real, we can separate the oscillating part of the field, $E_0e^{i(qx+k'z-\omega't)}$, from its amplitude,
\be
|E(x,z;t)|=|E_0| e^{-(k''z-\omega'' t)}\,,
\label{Exzt}
\ee
which is either exponentially decaying or exponentially growing in time and space.
Using the fact that $\sigma'>0$, which is equivalent to the positive imaginary part of the permittivity, we find from \Eq{k_TE} that $k'<0$, which physically corresponds to a plane wave propagating in vacuum towards the graphene sheet. The in-plane wave number $q$ is however much larger, $q\gg|k'|$, so that the direction of the electromagnetic wave is almost parallel to the sheet. At the same time, the $z$-coordinate of any point sitting on the wavefront and moving together with it can be approximately described as a function of time by $z(t)=z(0)+k't/q$. Substituting this into \Eq{Exzt}, we find the amplitude of the field at the selected point on the wavefront to be
\be
A(t)= A(0)e^{-k''z(t)+\omega'' t}= \tilde{A}(0)e^{-\gamma t}\,,
\ee
where we have introduced a temporal decay rate
\be
\gamma= \frac{k''k'}{q}-\omega''\,.
\label{Eq:gamma}
\ee
Now, substituting here the dispersion for the TE mode, given by \Eqs{w_TE}{k_TE}, we find
\be
\gamma_{\rm TE}=0\,,
\label{gamTE}
\ee
which implies that the TE mode has no losses in reality. In contrast, for the Drude-like TM mode, the decay rate found from \Eq{Eq:gamma} and the approximation \Eqs{w_TM}{k_TM} is given by
\be
\gamma_{\rm TM}=\sigma'(q-\sqrt{\omega_0q})/2>0\,,
\ee
which implies an absorption, since the amplitude $A(t)$ of the wave front decays with time in this case. However, for a Drude conductivity without damping, we obtain $\gamma_{\rm TM}=0$, since the real part of the conductivity $\sigma'$ vanishes, again implying that there are no losses in the system. The same is true for the graphene conductivity at zero temperature, since in this case $\sigma'=0$ for $\Omega<2$. We would like to emphasize, however, that \Eq{gamTE} is obtained for the TE mode at a non-zero temperature when the optical losses are present in the conductivity, since $\sigma'>0$ (equivalent to $\text{Im}\,\varepsilon>0$).

We see that the result for the TE mode, \Eq{gamTE}, is the same below and above the lower threshold at $\omtelow$, and the TE mode demonstrates a fully physical behaviour from the energy conservation viewpoint on both sides of the threshold, as discussed above. Since below the threshold, the TE mode has, without any gain, an exponential growth with time, albeit at a very small rate compared to the mode frequency, this mode possesses a unique property, which has never been reported in the literature, to the best of our knowledge.

\begin{figure}[t]
\centering
\includegraphics[width=\columnwidth]{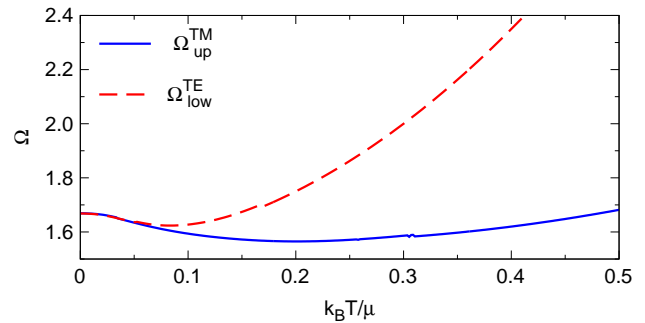}
\caption{Lower threshold frequency $\omtelow$ for the TE mode (red dashed line), and upper real frequency limit $\omtmup$ for the TM mode (blue solid line), as functions of the normalized temperature $(\mu\beta)^{-1}=k_BT/\mu$.}
\label{fig:O_0}
\end{figure}

\subsection{Temperature dependence of the threshold frequencies}
\label{Sec:temperature}

Finally, we study the temperature dependence of the TE threshold frequency, $\omtelow$. As discussed earlier in relation to \Eqs{w_TE}{k_TE},  this frequency threshold corresponds to a simultaneously change of sign of  the imaginary part of the conductivity and the mode frequency. The red dashed line in Fig.\,\ref{fig:O_0} showing the threshold frequency $\omtelow$ as function of temperate is thus a solution of the equation  $\text{Im}\,\sigma(\omega)=0$.  Figure~\ref{fig:O_0} clearly shows that the threshold frequency has a minimum, taking the value of $\omtelow=1.6225$ at $1/\mu\beta=0.0824$, which is the result of a trade-off of the two terms in the graphene conductivity, corresponding to the intraband and interband transitions.
More details on the properties of the TE mode near the lower threshold are provided in Appendix~\ref{range_of_sols}, where an analytic equation determining its value at zero temperature ($\omtelow=1.667$) is derived.

Interestingly, the TM modes has an upper threshold $\omtmup$ which  coincides with $\omtelow$ at zero temperature but deviates for non-zero temperatures. It is also shown in Fig.\,\ref{fig:O_0}, by a blue solid curve. This threshold, however, has a different physical meaning as it plays the role of an asymptote for the TM mode frequency  in the limit  of $q\to\infty$.  It can be seen that in this limit, \Eq{tm_secular} simplifies to $\sigma(\omega)=0$. Clearly, at zero temperature and frequencies below $\omteup=2$, this coincides with the equation $\text{Im}\,\sigma(\omega)=0$ for $\omtelow$, as can be seen also from \Fig{fig:condcomplex}\,(a) demonstrating that  $\text{Re}\,\sigma(\omega)=0$ in this frequency range.

\section{Conclusions}
\label{Sec:summary}
In summary, we have numerically calculated the complex-frequency dispersion of the surface plasmon polariton (SPP) modes in a homogeneous graphene layer both in transverse-magnetic (TM) and transverse-electric (TE) polarizations of light. We have further developed a simple analytic approximation which agrees well with the numerically exact solution of the secular equations for the modes in both polarizations.

We have shown that the TM SPP mode is determined by the Drude-like intraband part of the optical conductivity, demonstrating a square-root dispersion of the mode frequency with respect to the propagation  wave number. In this polarization, the temporal decay of the electromagnetic field, which is given by the imaginary part of the mode frequency, monotonously increases with temperature. The TE SPP mode is in turn determined by both the intraband and the interband part, the latter being crucial for its existence. Unlike the TM mode, its dispersion is close to the light line, and the temporal decay demonstrates a nontrivial dependence on temperature and the propagation wave number. We have observed, in particular, that at finite temperature and chemical potential the TE mode exists above the upper threshold
for the normalized frequency at $\Omega=2$, posed by an asymptotic behaviour of the dispersion at zero temperature. The temporal decay rate of the TE mode increases below the threshold, and decreases otherwise as the temperature increases. This is explained by considering occupation of electronic energy bands in graphene near the K-point at different temperatures.

We have also proven that the TE mode exists both above and below the lower threshold $\Omega=\omtelow$ (taking the value of $\omtelow=1.667$ at zero temperature) and have studied its behavior near the threshold. This threshold is caused by a change of the sign of the imaginary part of the graphene conductivity which in turn causes a simultaneous change of the sign of the imaginary part of the SPP complex eigenfrequency and the normal component of the wave number. By investigating its spatial and temporal evolution, we have shown that the TE SPP in graphene presents a unique optical mode, as it can have below the threshold a positive imaginary part of the eigenfrequency without introducing gain into the material. Furthermore, we have  demonstrated that in spite of the positive real part of the conductivity implying the positive imaginary part of the permittivity and hence an absorption, the TE SPP mode propagates along the graphene sheet without losses even at non-zero temperatures. This is correct at least up to second order in the conductivity.

\appendix

\section{Intraband conductivity of graphene}
\label{App:Intraband}

Following~\cite{Gusynin2006,Falkovsky2007,Falkovsky2007,Mikhailov2007}, one can derive  the two-dimensional (2D) optical conductivity of a homogeneous graphene sheet by using the Kubo formula~\cite{Bruus2004,Stauber2008,Hanson2008,Mahan2012}, namely, by expanding the thermodynamic average of the current to first order in the amplitude of an external electric field.
In the long wavelength limit, i.e. for a small light wave number compared to that of the electron, one may neglect effects of the spatial dispersion. The intraband conductivity is then given by~\cite{Mikhailov2007}
\begin{align}
    \sigma_{\alpha \beta}^\text{intra}(\omega) = -\frac{ie^2g_sg_v}{\hbar^2\omega S}\sum_{\mathbf{k}l}
    \frac{\partial E_{\mathbf{k}l}}{\partial{k}_\alpha}
    \frac{\partial f(E_{\mathbf{k}l})}{\partial E_{\mathbf{k}l}}
    \frac{\partial E_{\mathbf{k}l}}{\partial {k}_\beta}\,,
    \label{sigma_intra_xx}
\end{align}
where $g_s=2$ and $g_v=2$ are, respectively, the spin and valley degeneracies, $S$ is the sample area,
\be
E_{\mathbf{k}l} = (-1)^l\hbar V k =  (-1)^l\hbar V \sqrt{k_x^2+k_y^2}
\label{disp}
\ee
is the electron dispersion in graphene, with $l=1$ and 2,
\be
f(E) = \frac{1}{1+e^{\beta(E - \mu)}}
\ee
is the Fermi-Dirac distribution function, and $V$ is the electron Fermi velocity.
Since we neglect the spatial dispersion, the conductivity tensor has a diagonal symmetric form, $\sigma_{xx}=\sigma_{yy}=\sigma$ with $\sigma_{xy}=\sigma_{yx}=0$.
The expressions for the partial derivatives are given by
\be
    \frac{\partial E_{\mathbf{k}l}}{\partial k_\alpha}= (-1)^l\hbar V\frac{k_\alpha}{k}
\label{der}
\ee
which we substitute into \eqref{sigma_intra_xx} to obtain
\bea
    \sigma^\text{intra}(\omega) &=& \frac{-4ie^2}{\hbar^2 \omega (2\pi)^2}\int\!\! \int dk_x dk_y
    \hbar^2V^2\frac{k_x^2}{k^2} \sum_{l=1}^2 \frac{\partial f(E_{\mathbf{k}l})}{\partial E_{\mathbf{k}l}}
    \nonumber
    \\
    &=& -\frac{ie^2}{\pi \hbar^2 \omega} \int_0^\infty \frac{\partial g(E)}{\partial E} EdE\,,
\eea
after introducing $ g(E)=f(E)-f(-E)$ and performing integration over the angle in polar coordinates.
The last integral can be evaluated analytically using integration by parts,
\be
\int_0^\infty \frac{\partial g(E)}{\partial E}EdE=\left.E g(E)\right|_0^\infty-\int_0^\infty g(E)dE
=\left.G(E)\right|_0^\infty\,,
\ee
where
\be
G(E)=E [f(E)-f(-E)]- \frac{1}{\beta} \ln [f(E)f(-E)]\,,
\ee
which follows from the fact that for the Fermi function,
\be
\frac{\partial f(E)}{\partial E}=\beta f( E)[f( E)-1]\,.
\ee
Applying the limits of integration, we obtain
\bea
G(0)&=&\frac{1}{\beta} \ln (1+e^{-\beta\mu})^2\,,\\
G(\infty)&=&\lim_{E\to \infty} \left[ - E +\frac{1}{\beta} \ln e^{\beta(E-\mu)}\right]=-\mu\,,
\eea
so that
\be
G(\infty)-G(0)= -\frac{1}{\beta} \ln \left[2+2\cosh (\beta\mu)\right]
\ee
and finally
\be
\sigma^\text{intra}(\omega) = \frac{ie^2 \ln[2+2\cosh (\beta\mu)]}{\pi \hbar^2 \beta\omega} \,.
\ee
At zero temperature, $\mu\beta \to\infty$, the intraband conductivity simplifies to
\be
    \sigma^\text{intra} = \frac{ie^2\mu}{\pi\hbar^2 \omega}\,.
\ee

\section{Interband conductivity of graphene}
\label{App:Interband}

The interband conductivity is derived in a similar way, i.e. again using the Kubo formula, which leads in the long wavelength limit to the following expression~\cite{Mikhailov2007}:
\begin{align}
    \sigma_{\alpha \beta}^\text{intra}(\omega) = &\frac{ie^2\hbar g_sg_v}{\omega S}\sum_{\mathbf{k},l\neq l'}
    \frac{f(E_{\mathbf{k}l'})-f(E_{\mathbf{k}l}) }{E_{\mathbf{k}l'}-E_{\mathbf{k}l}-\hbar(\omega+i0_+)}
    \nonumber\\
    &\times
  \frac{1}{E_{\mathbf{k}l'}-E_{\mathbf{k}l}}
  \langle \mathbf{k}l|\hat{v}_\alpha |\mathbf{k}l'\rangle
   \langle \mathbf{k}l'|\hat{v}_\beta |\mathbf{k}l\rangle\,,
    \label{sigma_inter_xx}
\end{align}
where $0_+$ is a positive infinitesimal, and
\be
\hat{v}_\alpha =V\hat{\sigma}_\alpha
\ee
with $\hat{\sigma}_\alpha$ being the Pauli matrix. $|\mathbf{k}l\rangle$ are the eigenstates of the electronic Hamiltonian near the $K$-point in the Brillouin zone,
\be
\hat{H}=V\hat{\mbox{\boldmath{$\sigma$}}}\cdot\hat{\mathbf{p}}\,,
\ee
corresponding to its eigenvalues \Eq{disp} and
having the following explicit form:
\be
|\mathbf{k}l\rangle = \frac{k}{\sqrt{2}}
\left(
\begin{array}{cc}
k_x-ik_y \\ (-1)^l k
\end{array}
\right).
\ee
The matrix elements in \Eq{sigma_inter_xx} then take the form
\be
\langle \mathbf{k}1|\hat{v}_x |\mathbf{k}2\rangle=\frac{i k_y}{k}\,,\ \ \ \
\langle \mathbf{k}1|\hat{v}_y |\mathbf{k}2\rangle=-\frac{i k_x}{k}\,.
\ee
Again using the symmetry of the conductivity tensor in the absence of the spatial dispersion, we obtain with the help of \Eq{disp} and after integration over the angle in polar coordinates
\bea
    \sigma^\text{inter}(\omega) &=& \frac{4ie^2\hbar}{(2\pi)^2}\int\!\! \int dk_x dk_y
    V^2\frac{k_y^2}{k^2} \sum_{l=1}^2
\frac{1}{E_{\mathbf{k}l'}-E_{\mathbf{k}l}}
    \nonumber\\
    &&\times
\frac{f(E_{\mathbf{k}l'})-f(E_{\mathbf{k}l}) }{E_{\mathbf{k}l'}-E_{\mathbf{k}l}-\hbar(\omega+i0_+)}
   \nonumber\\
&=& \frac{ie^2}{2\pi \hbar } \int_0^\infty dE [f(-E)-f(E)]
\\
    &&\times \left[\frac{1}{\hbar(\omega+i0_+)-2E}+\frac{1}{\hbar(\omega+i0_+)+2E}\right]\,.
\nonumber
  \eea

\section{SPP modes of a thin conducting sheet in vacuum}
\label{App:modes}
To derive the secular equations~(\ref{tm_secular}) and (\ref{te_secular}) for the SPP modes in a graphene layer, let us consider a model of an infinitely thin sheet with 2D optical conductivity $\sigma(\omega)$, placed at $z=0$. Choosing $x$ as the propagation direction of light, so that the parallel components of the wave number are $k_x=p$ and $k_y=0$, Maxwell's equations are split into two blocks of first-order partial differential equations, separating TM and TE polarizations~\cite{NealePRB20}:
\begin{align}
    \text{TM: }
    \begin{pmatrix}
    \omega\mu(\omega;z) & -\partial_z & iq\\
    \partial_z & \omega\varepsilon(\omega;z) & 0\\
    -iq & 0 & \omega\varepsilon(\omega;z)
    \end{pmatrix}\begin{pmatrix}
    iH_y\\E_x\\E_z
    \end{pmatrix}=0\,, \label{tmblock}\\
    \text{TE: }
    \begin{pmatrix}
    \omega\varepsilon(\omega;z) & -\partial_z & iq\\
    \partial_z & \omega\mu(\omega;z) & 0\\
    -iq & 0 & \omega\mu(\omega;z)
    \end{pmatrix}\begin{pmatrix}
    E_y \\iH_x\\iH_z
    \end{pmatrix}=0\,, \label{teblock}
\end{align}
where the speed of light in vacuum is taken $c=1$ for brevity, $\partial_z\equiv\partial/\partial z$,
$\varepsilon(\omega;z)$ and $\mu(\omega;z)$ are, respectively, the frequency and spatially dependent permittivity and permeability, $\bm{E}=A_0(E_x,E_y,E_z)$ and $\bm{H}=A_0(H_x,H_y,H_z)$ are, respectively, the electric and magnetic fields, $\omega$ is the light frequency, and $A_0(x,t)=e^{i(qx-\omega t)}$ is a common factor representing the temporal behaviour and the spatial dependence of the fields in the propagation direction. Clearly, \Eqs{tmblock}{teblock} can be obtained from each other by simultaneous swapping $\varepsilon\leftrightarrow\mu$ and $\bm{E}\leftrightarrow i\bm{H}$~\cite{NealePRB20}.

Let us further assume that
\begin{align}
    \mu(\omega;z)&=1\,,\label{app31}\\
    \varepsilon(\omega;z)&=1+\chi(\omega) \delta(z)\,,\label{app32}
\end{align} where
\be
\chi(\omega)=\frac{2i\sigma(\omega)}{\omega}
\label{gamma}
\ee
is a 2D susceptibility
and $\sigma(\omega)$ is the 2D electrical conductivity of the graphene sheet,
\be
\sigma(\omega)=2\pi[\sigma^\text{intra}(\omega)+\sigma^\text{inter}(\omega)]\,,
\label{sigma}
\ee
consisting of the intraband and interband components calculated in Appendices~\ref{App:Intraband} and \ref{App:Interband}. Note that the factor of 2 affecting the definition of $\sigma$ is introduced in \Eq{gamma} for convenience.

\subsection{TM polarization}\label{tm_appendix}
The set of equations~(\ref{tmblock}) for TM polarization then simplifies to
\begin{align}
    i\omega H_y-\partial_zE_x+iqE_z&=0\,,\label{tm71}\\
    i\partial_zH_y+\omega\varepsilon E_x&=0\label{tm72}\,,\\
    qH_y+\omega\varepsilon E_z&=0\label{tm73}\,.
\end{align}
For $z\neq 0$, we find the wave equation for $H_y$ and relations between the field components:
\begin{align}
    \partial_z^2&H_y+(\omega^2-q^2)H_y=0,\,\label{tm81}\\
    &E_x=-\frac{i}{\omega}\partial_zH_y,\,\label{tm82}\\
    &E_z=-\frac{q}{\omega}H_y\,.\label{tm83}
\end{align}
To include the $z=0$ point, we use \Eq{app32} and integrate \Eqsss{tm71}{tm72}{tm73} over an infinitesimal interval including $z=0$. Then we obtain
\begin{align}
    E_x(0_+)-E_x(0_-)&=0\label{tm91}\,, \\
    -H_y(0_+)+H_y(0_-)+i\omega\chi E_x(0)&=0\label{tm92} \,,\\
    \omega \chi E_z(0) &= 0\,,\label{tm93}
\end{align}
where $0_+$ ($0_-$) is a positive (negative) infinitesimal. From the above equations \eqref{tm81}\,--\,\eqref{tm93} we find
\begin{align}
    H_y&=A\,\text{sgn}(z)e^{ik|z|}\,, \\
    E_x&=A\frac{k}{\omega}e^{ik|z|} \,,\\
    E_z&=-A\frac{q}{\omega}\text{sgn}(z)e^{ik|z|}\,,
\end{align} after applying outgoing or incoming wave boundary conditions to the electro-magnetic field. Here $A$ is a normalization constant,
\begin{align}
    \text{sgn}(z)=\begin{cases} 1\,,& z>0\\0\,,& z=0\\-1\,,& z<0\,,
    \end{cases}
\end{align}
and $k$ is the normal component of the wave number in vacuum satisfying the light dispersion
\begin{align}
    \omega^2=k^2+q^2\,.
\label{omega-k}
\end{align}
Finally, using \Eq{tm92}, we obtain a secular equation for the SPP mode:
\be
    ik\chi=2\,,
\ee
which can be written more explicitly, using \Eq{gamma},  as
\be
    k{\sigma(\omega)}+\omega=0\,,
\label{TM-sec}
\ee
identical to \Eq{tm_secular}.

To obtain an approximate analytic solution to \Eq{TM-sec}, let us first consider the limiting case of very small frequencies when the conductivity is dominated by intraband transitions. This results in the standard SPP mode of an undamped Drude metal sheet. In fact, in this case
\be
\sigma(\omega)\approx  2\pi\sigma^{\rm intra}(\omega)= \frac{i\omega_0}{\omega}\,,
\label{Drude}
\ee
where
\be
\omega_0= 2\alpha \frac{\ln(2+2\cosh{\mu\beta})}{\hbar\beta}\,,
\ee
see \Eq{sigma_formula} (at zero temperature $\omega_0$ simplifies to just $\omega_0= 2\alpha \mu/\hbar$). Then the solution of \Eq{TM-sec} takes the form:
\be
\bar{k}=i\varkappa\,, \ \ \ \bar{\omega}= \sqrt{\varkappa \omega_0}\,,
\label{SPP-Drude}
\ee
where
\be
\varkappa=\sqrt{q^2+\omega_0^2/4}-\omega_0/2\,.
\ee

Now, taking the full 2D conductivity of graphene, \Eq{sigma},
and treating $\Delta\sigma =2\pi\sigma^{\rm inter}$ as a correction to the Drude conductivity, \Eq{Drude}, results in a refinement of the SPP mode wave number and frequency, $ k=\bar{k}+\Delta k$ and $\omega=\bar{\omega}+\Delta \omega$. We find in particular from \Eqs{omega-k}{TM-sec}
\be
-(i\varkappa+\Delta k)(i\omega_0+\omega \Delta\sigma) = \omega^2=(i\varkappa+\Delta k)^2+q^2\,,
\ee
and keeping in the above equation only terms linear in $\Delta k$ and $\Delta \sigma$, obtain
\be
\Delta k\approx-\frac{\varkappa \bar{\omega}}{2\varkappa+\omega_0} \Delta \sigma\,.
\ee
Then, from \Eq{omega-k} we find
\be
\bar{\omega}\Delta\omega\approx \bar{k}\Delta k\,,
\ee
which results in
\be
\Delta \omega\approx-i \frac{\varkappa^2 \bar{\omega}}{2\varkappa+\omega_0} \Delta \sigma\,.
\ee

Finally, in the limit $q\gg\omega_0$ we obtain
\bea
\varkappa &\approx& q\,,\ \ \ \ \ \ \ \ \ \ \ \ \ \ \ \bar{\omega}\approx\sqrt{q\omega_0}\,,\\
\omega' &\approx& \bar{\omega}\,,\ \ \ \ \ \ \ \ \ \ \ \ \ \ \ \omega'' \approx -\frac{q}{2}\sigma'(\bar{\omega})\,,\\
k' &\approx& -\frac{\bar{\omega}}{2}\sigma'(\bar{\omega})\,,\ \ \ \ k''\approx q\,,
\eea
separating the real and imaginary parts of $k=k'+ik''$, $\omega=\omega'+i\omega''$, and
$\sigma=\sigma'+i\sigma''$.

\subsection{TE polarization}\label{te_appendix}
The secular equation for TE polarization is obtained in a similar way. Using \Eqs{app31}{app32}, the TE block given by \Eq{teblock} can be written as
\begin{align}
    \omega\varepsilon E_y -i\partial_zH_x-qH_z&=0\,,\label{app181}\\
    \partial_zE_y+i\omega H_x&=0\,,\label{app182}\\
    -qE_y+\omega H_z&=0\,.\label{app183}
\end{align}
For $z\neq 0$, we obtain
\begin{align}
    \partial_z^2&E_y+(\omega^2-q^2)E_y=0\,,\label{app191}\\
    &H_x=\frac{i}{\omega}\partial_zE_y\,,\label{app192}\\
    &H_z=\frac{q}{\omega}E_y\,,\label{app193}
\end{align}
and for $z=0$, we integrate \Eq{app181} around this point, obtaining
\begin{align}
    H_x(0_+)-H_x(0_-)=-i\omega\chi E_y(0)\,,
    \label{app20}
\end{align}
at the same time having both $E_y$ and $H_z$ continuous across $z=0$. A solution satisfying outgoing or incoming wave boundary conditions then takes the form:
\begin{align}
    E_y(z)&=Ae^{ik|z|},\label{app211}\\
    H_x(z)&=-A\frac{k}{\omega}\text{sgn}(z)e^{ik|z|},\label{app212}\\
    H_z(z)&=\frac{q}{\omega}Ae^{ik|z|},\label{app213}
\end{align}
and \Eq{app20} provides a secular equation for the SPP mode:
\be
    2ik=-\omega^2\chi\,,
\label{TE-sec}
\ee
which can be written more explicitly, using \Eq{gamma},  as
\be
    k+\omega\sigma(\omega) = 0\,,
\ee
identical to \Eq{te_secular}.

\begin{figure}[t]
\centering
\includegraphics[width=\columnwidth]{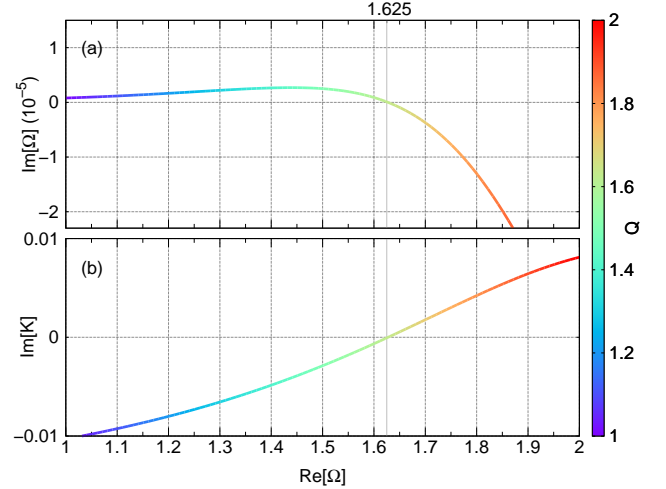}
\caption{Imaginary part of (a) the TE mode frequency $\Omega$ and (b) the normal component of the light wave number $K$, as functions of the real part of $\Omega$, calculated for $\mu\beta=10$ for changing the in-plane light wave number $Q$ given by the color code.}
\label{te_10_classify}
\end{figure}
To obtain an approximate analytic solution of \Eq{TE-sec}, let us use the fact that $|\sigma(\omega)|\ll 1$ if the frequency is not too small. This is due to the fact that $\sigma$ is proportional to the fine-structure constant $\alpha$ which is a small number. Combining \Eqs{omega-k}{TE-sec}, we obtain
\be
\omega\approx q+\frac{q}{2}\sigma^2(q)\,, \ \ \ \ \ k\approx -q\sigma(q)\,.
\ee
Extracting the real and imaginary parts we then find
\bea
\omega' &\approx& q\,,\ \ \ \ \ \ \ \ \ \ \ \ \ \omega'' \approx q\sigma'(q)\sigma''(q)\,,\\
k' &\approx& -q\sigma'(q)\,,\ \ \ \ k''\approx -q\sigma''(q)\,.
\eea
Interestingly, $k'<0$,  since $\sigma'(\omega)> 0 $, at least for a real frequency -- the same as in TM polarization. This implies that the light in the SPP mode propagates both along and towards the graphene sheet. At the same time, the amplitude of the wave exponentially decreases (increases) with distance from the sheet when $ \sigma''(\omega)<0 $ ($ \sigma''(\omega)>0 $). We also see that the sign of $k''$ and $\omega''$ changes simultaneously at the lower threshold frequency as we discuss in detail in \Sec{Sec:modes}.

\begin{figure}[t]
\centering
\includegraphics[width=\columnwidth]{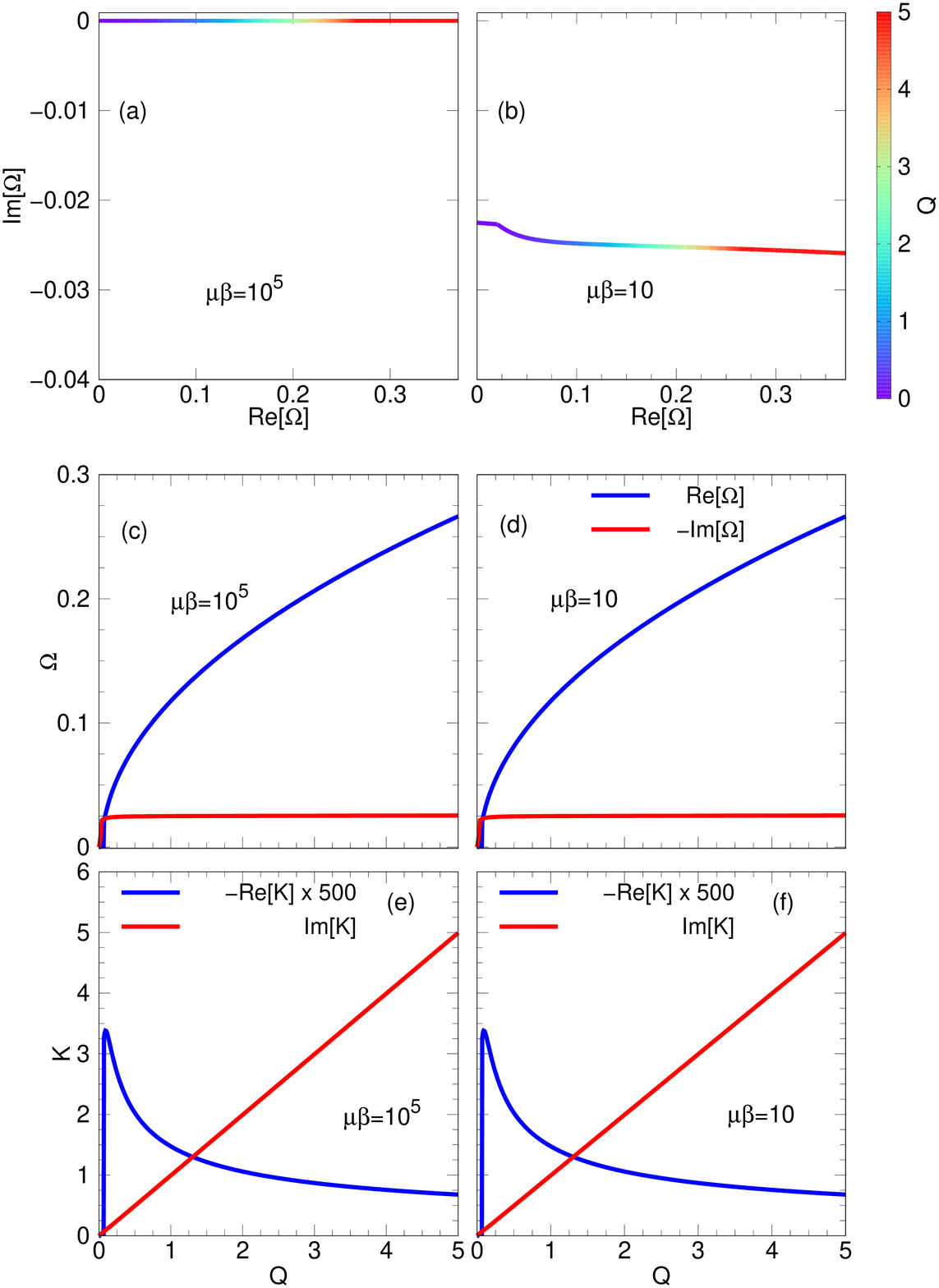}
\caption{As \Fig{tm_plane} but for $\Gamma=0.05$ and without using the analytic approximation
} \label{gamma_tm_plane}
\end{figure}

\begin{figure}[t]
\centering
\includegraphics[width=\columnwidth]{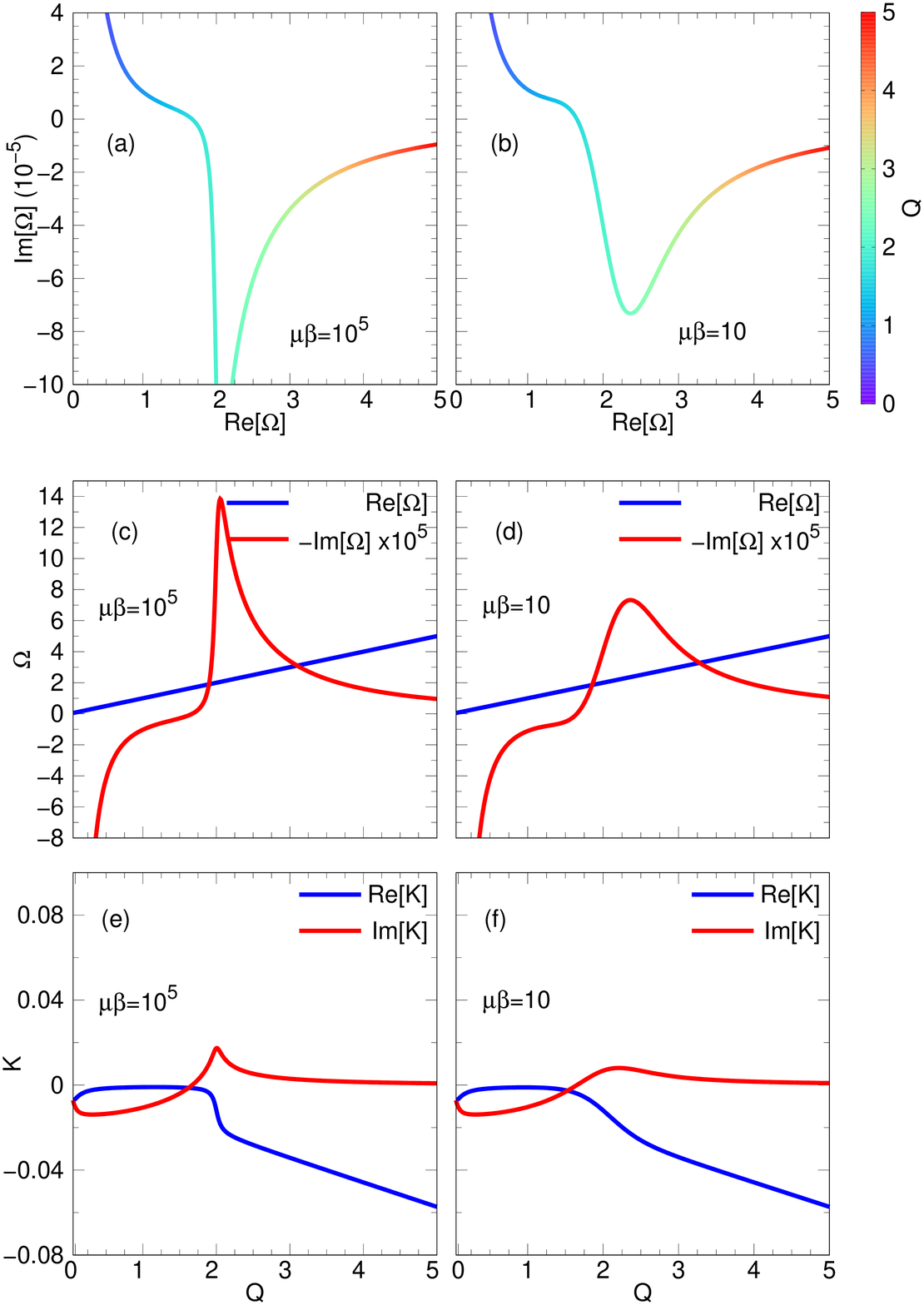}
\caption{
As \Fig{te_plane} but for $\Gamma=0.05$ and without using the analytic approximation
}
\label{gamma_te_plane}
\end{figure}

\section{Lower threshold frequency for TE polarization}\label{range_of_sols}
In this appendix, we discuss in more detail the condition 
for the lower threshold frequency $\omtelow$ of TE mode and derive an equation determining the threshold value of $\omtelow\approx1.667$ at zero temperature.

Let us first note that, in deriving \Eq{te_secular}, the electromagnetic field of the wave coupled to charge oscillations is proportional to
\begin{align}
    e^{ik|z|}\,,
\end{align} where $k=\sqrt{\omega^2-q^2}$, $q$ is the in-plane wave number in the direction of travel, and $z$ is distance to the graphene sheet. For a bounded solution, we therefore assert that  
\begin{align}
    \text{Im}\,k>0\,.\label{bounded_assertion}
\end{align}
In the opposite case, $\text{Im}\,k<0$, the electromagnetic field would grow exponentially away from the graphene layer, and the threshold frequency is define as a value at $\text{Im}\,k=0$. As an example for  finite temperature $\mu\beta=10$, we see in \Fig{te_10_classify} that the signs of both $\text{Im}\,k$ and  $\text{Im}\,\omega$ change simultaneously at the threshold frequency of $\omtelow\approx 1.625$. This is in agreement with the analytic approximation given by \Eqs{w_TE}{k_TE}.

The value of the threshold frequency depends on the temperature (and the chemical potential) as demonstrated by \Fig{fig:O_0}. It satisfies a general equation
\be
\text{Im}\,\sigma(\omega)=0\,,
\label{Eq:thr}
\ee
which can be easily obtained from \Eq{te_secular} by using the fact that $\text{Im}\,k=0$ at the threshold, and hence the mode frequency $\omega$ is real. Then taking the imaginary part of \Eq{te_secular} results in $k''+\omega\sigma''(\omega)=0$ which in turn gives \Eq{Eq:thr}.




Finally, we consider the limit of zero temperature (and $\Gamma=0$), in which case the imaginary part of the conductivity takes the form
\be
\text{Im}\,\sigma= \alpha\left(\frac{2}{\Omega}+\frac{1}{2}\ln\frac{2-\Omega}{2+\Omega}\right)
\ee
as it follows from \Eq{sigma_formula}, after performing the analytic integration in the interband part. Applying the threshold condition  \Eq{Eq:thr} then leads to
\begin{align}
    2+\omtelow =(2-\omtelow)\exp\left(\frac{4}{\omtelow}\right)\,.
\end{align}
The numerical solution of this equation gives $\omtelow\approx 1.667$.
Figure~\ref{fig:O_0} shows that $\omtelow$ approaches this value in the limit $T\to0$.

\section{SPP modes with non-zero damping}
\label{Sec:gamma}

In this appendix, we show the SPP dispersion relation with non-zero phenomenological damping for both TM and TE polarizations.
Figures~\ref{gamma_tm_plane} (a-d) show the dispersion of SPP TM mode for $\Gamma=0.05$. Here we observe that, compared to zero damping, $\text{Im}\,\omega$ is mainly lifted up by $\Gamma/2$ throughout the range of $q$. Whereas $\text{Im}\,k$, shown as red curves in Figs.\,\ref{gamma_tm_plane} (e) and (f), is unchanged compared to the case of $\Gamma=0$ (\Fig{tm_plane}),  $\text{Re}\,k$, shown as blue curves in the same figures, changes its behaviour and its values significantly,  increasing fast around $q=0$ and then gradually decreasing with $q$.

In contrast, the SPP TE mode shown in Figs.\,\ref{gamma_te_plane} (a-d),  demonstrates more changes in $\text{Im}\,\omega$.
In comparison with the dispersion for zero $\Gamma$ (\Fig{te_plane}),  the SPP TE mode has now smoother dependencies of $\text{Im}\,\omega$  and  $k$.
$\text{Im}\,k$, shown in Figs.\,\ref{gamma_te_plane} (e) and (f) starts from a negative value at $q=0$.
The TE SPP mode frequency still has a positive imaginary part in the range below the threshold frequency $\omtelow$, similar to what we have seen for $\Gamma=0$.

\bibliography{MyCollection}

\end{document}